\documentclass[aps,prb,twocolumn]{revtex4}   

\usepackage{amsmath}    
\usepackage{graphicx}   

\begin{document}

\newcommand{\ket}[1]{|#1\rangle}
\newcommand{\bra}[1]{\langle#1|}

\newcommand{\0}{|0\rangle}
\newcommand{\1}{|1\rangle}
\newcommand{\+}{|+\rangle}
\newcommand{\m}{|-\rangle}
\newcommand{\Cbas}{\{\0,\1 \}}
\newcommand{\Hbas}{\{\+,\m \}}

\newcounter{therule}
\newenvironment{Rule}[1]
{\stepcounter{therule}
\vspace{4ex} \par\noindent \textbf{\large{Rule \Roman{therule}: #1.}}
\vspace{3ex}\begin{itshape}\par\noindent}%
{\end{itshape}

\vspace{3ex}}

\newcounter{thecorollary}
\newenvironment{Corollary}[1]
{\stepcounter{thecorollary}
\vspace{3ex} \par\noindent \textbf{Corollary \Roman{therule}.\Alph{thecorollary}:}\vspace{1ex}\begin{itshape}\par\noindent}%
{\end{itshape}\vspace{3ex}}

\title{Equivalent Quantum Circuits}
\author{Juan Carlos Garcia-Escartin}

 \affiliation{Universidad de Valladolid, Dpto. Teor\'ia de la Se\~{n}al e Ing. Telem\'atica, Paseo Bel\'en n$^o$ 15, 47011 Valladolid, Spain}
 \email{juagar@tel.uva.es}   
\author{Pedro Chamorro-Posada}
\affiliation{Universidad de Valladolid, Dpto. Teor\'ia de la Se\~{n}al e Ing. Telem\'atica, Paseo Bel\'en n$^o$ 15, 47011 Valladolid, Spain}
\date{\today}

\begin{abstract}
Quantum algorithms and protocols are often presented as quantum circuits for a better understanding. We give a list of equivalence rules which can help in the analysis and design of quantum circuits. As example applications we study quantum teleportation and dense coding protocols in terms of a simple XOR swapping circuit and give an intuitive picture of a basic gate teleportation circuit.
\end{abstract}

\maketitle

\section{The circuit point of view}
\label{circuit}
Quantum communication and computation study information transmission and processing as physical phenomena that follow the laws of quantum mechanics. Considering quantum mechanics introduces new possibilities like private communication with quantum cryptography or efficient factoring algorithms among others \cite{NC00,Mer07}. 

Most quantum information protocols and algorithms can be explained as a sequence of transformations applied to a known initial state and a final measurement stage. The intermediate evolution is usually the key to the procedure. This state evolution can be studied from different perspectives. In this paper we take the point of view of quantum circuits.

In electronic and electrical engineering the circuit representation is routinely used to study classical electronic circuits. Circuit equivalences help to analyse complex processing blocks or to explain different logical operations. We present some quantum circuit equivalences which can play a similar role in quantum computation. Many of these equivalence rules have been used before in quantum circuit analysis \cite{ZLC00,Mer01,Mer02,MDM08}. We compile some of the most useful and provide new derivations. 

\begin{figure}[ht!]
\centering
\includegraphics[scale=1]{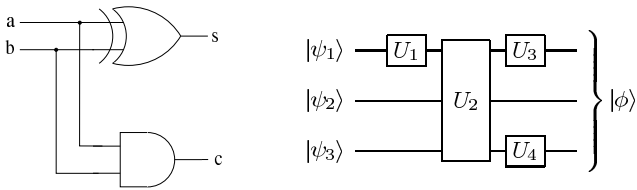}
\caption{Classical logical circuit (left) and quantum logical circuit (right).\label{cqcircuits}} 
\end{figure}

Most of the conventions for quantum circuit representation are taken from classical circuits. There are some wires (usually represented as lines) which carry the signals (states) to different points of the circuit. The basic operations are represented as \emph{gates}. Figure $\ref{cqcircuits}$ shows two example circuits, a classical circuit (a half adder), and a generic quantum circuit. We follow the usual convention of a state going from left to right, like an electrical signal traversing the electronic elements. 

The paper starts with a brief review of quantum computation (Section \ref{fundamentals}) which readers already familiar with the basics can skip. In Section \ref{gates} we set down the notation and define all the gates we will employ in the rest of the paper. Section \ref{rules} gives the list of transformation rules. Sections \ref{statetransf} and \ref{bellstates} present some simple quantum computation blocks which appear in many applications. Finally, Section \ref{examples} goes through some examples in which the given rules can be used to understand basic quantum protocols.

\section{Fundamentals of quantum computation}
\label{fundamentals}

\subsection{The quantum information unit: the qubit}
Information can be represented in many possible formats. In this paper we consider discrete quantum information units, the qubits. 

\begin{figure}[ht!]
\centering
\includegraphics{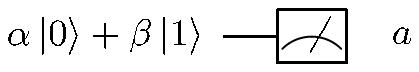}
\caption{Measurement in a quantum circuit. The binary outcome of the measurement can be associated to a classical bit $a$, which takes value 0 if the state $\ket{0}$ is found (with probability $|\alpha|^2$) and takes value 1 when the state is $\ket{1}$ (with probability $|\beta|^2$).\label{measure}} 
\end{figure}

In classical computers, information is stored in bits that can be either 0 or 1. Their quantum counterpart, the qubits, are binary quantum information units that can exist in an arbitrary superposition of states of the form 
\begin{equation}
\ket{\psi}=\alpha \ket{0}+\beta \ket{1},
\end{equation}
 where $|\alpha|^2$ and $|\beta|^2$, such that $|\alpha|^2+|\beta|^2=1$, are the respective probabilities of finding $\ket{0}$ and $\ket{1}$ after a measurement in the $\{\0,\1\}$ basis. Figure \ref{measure} shows the circuit representation of a measurement. A measurement on a qubit has two possible outcomes which can be associated to the binary value of a classical bit. 

\subsection{Multiple qubits: tensor product and entanglement}
A system with $n$ qubits can be expressed as a complex vector in a Hilbert space of dimension $2^n$. If the qubits are not correlated, the composite system comes from taking the tensor product of the individual state of the $n$ qubits (see \cite{Mer07} for a good operational description). Tensor products are indicated by the symbol $\otimes$. Two states $\ket{\psi_1}$ and $\ket{\psi_2}$ have a joint state $\ket{\psi}=\ket{\psi_1}\otimes\ket{\psi_2}$.

Any complex unit vector in our Hilbert space represents a valid quantum state. However, not all the valid states can be expressed as a collection of independent qubits. When the state can only be described as a whole, we say it is entangled. 

Entanglement is probably the most counterintuitive result from quantum mechanics and is an essential resource in quantum information. In Section \ref{examples} we will see that the striking properties of the quantum dense coding and teleportation protocols boil down to the capacity of sharing long distance correlations.
  
\subsection{Quantum gates}
Quantum evolution can be described with unitary operators $U$ acting on the quantum states. The physical system which provides such a unitary evolution is called a quantum gate. Evolution through unitary operators is indicated by multiplication. If we start from a state $\ket{\psi_0}$ and apply unitary operator $U_1$ to obtain state $\ket{\psi_1}$, this is written as $\ket{\psi_1}=U_1\ket{\psi_0}$. A sequence of operators $U_1$, $U_2$, $U_3$, $\ldots U_n$ applied in order of increasing index (first $U_1$, then $U_2$ and so on) is written from left to right $\ket{\psi_n}=U_nU_{n-1}\cdots U_2U_1\ket{\psi_0}$. The operator closest to the initial state $\ket{\psi_0}$ is the one which is applied first. This sequence is equivalent to a single operator $U=U_nU_{n-1}\cdots U_2U_1$. 

The order can be better understood from a simple representation in terms of linear algebra operators. For most quantum information purposes, we can imagine states from a space of dimension $N$ are complex column vectors with $N$ rows and the operators $U_i$ are $N\times N$ unitary matrices. Unitary operators (and matrices) do not commute in general. This is why the right order must be preserved. 

If we only act on some of the individual qubits, the effect on the joint system can be described using a tensor product of operators. We place an identity operator $I$ for all the positions in which there is no change. For instance, in the quantum circuit of Figure \ref{cqcircuits} the global evolution is given by $U=(U_3\otimes I \otimes U_4)U_2(U_1\otimes I \otimes I)$, where $I$ is the qubit identity. Notice the gate order. In the circuit representation gates are written from left to right. We imagine a quantum state travelling to the right which sees a series of different gates on its way out of the circuit.

\section{Basic quantum gates}
\label{gates}
Quantum gates are quantum analogous to the digital gates of electronic digital computers (such as AND, OR, NOT, XOR\ldots). A combination of quantum gates forms a quantum circuit. An $n$-qubit quantum gate can be defined as a system that performs a determined operation on $n$ input qubits so that for each input value there is a defined associated output. Superpositions of different input states will produce the corresponding superposition of output states. 

We will treat the operations of quantum computing from a circuital point of view. In the circuital model, gates are presented following the classical circuit convention: the input is drawn at the left and the circuit gates are presented in order of application from left to right. The input qubits travel through all these gates and, at the output of the circuit, at the right, they emerge transformed. 

Quantum gates can be grouped into families. We will be mostly concerned with three kinds of gates: single qubit gates, controlled gates and classically controlled gates. 

\subsection{Single qubit gates}
Unlike classical logic, which only admits the NOT operation for single bits, the complex nature of the probability amplitudes associated to each quantum logical value and the ability to form superpositions allow for a richer interplay in a single qubit.

In many applications we will only need three different single qubit gates: NOT, Hadamard and Z gates.

\subsubsection{X gate}
The NOT, or X, gate is the quantum generalization of the classical NOT gate and flips the value of the qubit it is acting on. After an X gate, $\0$ becomes $\1$ and $\1$ becomes $\0$. Usually, this is written as $X\ket{x}=\ket{x\oplus 1}$, where $\oplus$ is used to account for a XOR, or modulo 2 addition, operation (see Table \ref{XORtruth}). 

\begin{table}[h!]
\begin{center}
\begin{tabular}{| cc | c |}	
\hline
\multicolumn{3}{|c|}{\bf{XOR}}\\
\hline
\phantom{a}a\phantom{a}&\phantom{a}b\phantom{a}&\phantom{a}$a\oplus b$\phantom{a}\\
\hline
0&0&0\\
0&1&1\\
1&0&1\\
1&1&0\\
\hline
\end{tabular}
\caption{Truth table for the XOR logical operation.}
\label{XORtruth}
\end{center}
\end{table}

\subsubsection{H gate}
\label{Hgat}
The operation of the H gate can be seen from its effect on the states of the computational basis, $\Cbas$. These states are transformed into two orthogonal superpositions
\begin{equation}
H\ket{0}=\ket{+}=\frac{\ket{0}+\ket{1}}{\sqrt{2}}
\end{equation} 
and 
\begin{equation}
H\ket{1}=\ket{-}=\frac{\ket{0}-\ket{1}}{\sqrt{2}}.
\end{equation}
Superpositions of $\0$ and $\1$ result in the corresponding superpositions of $\+$ and $\m$. A compact way to express the operation is 
\begin{equation}
H\ket{x}=\frac{\ket{0}+(-1)^x\ket{1}}{\sqrt{2}}
\end{equation}
where $\ket{x}$ is a state from the computational basis ($x$ is either 0 or 1). 

This gate is its own inverse, as $H\ket{+}=\ket{0}$ and $H\ket{-}=\ket{1}$.

\subsubsection{Z gate}
The Z gate performs a sign shift when the value of the qubit is $\1$ and does nothing otherwise. The operation can be written as $Z\ket{x}=(-\!1)^x\ket{x}$. The Z gate belongs to a more general family of phase shift gates which introduce a $\Phi$ phase shift on the $\1$ state. The Z gate corresponds to a phase shift of $\Phi=\pi$. 

\begin{figure}[h]
\centering
\includegraphics{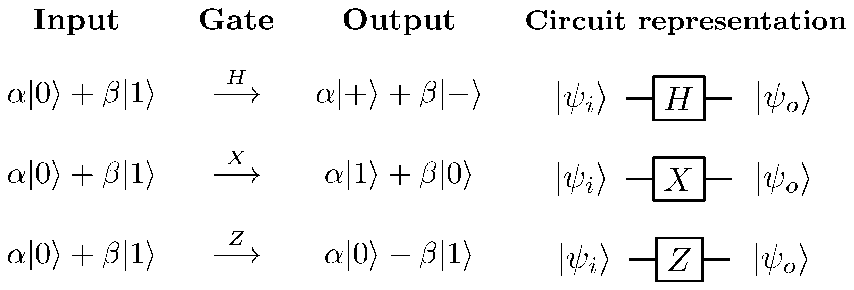}
\caption{Operation of selected one-qubit gates and their circuit representations.\label{singqubitoperation}} 
\end{figure}

Figure \ref{singqubitoperation} sums up the operation of each of these gates and provides the most usual circuit representation which will be used in later circuits. It is worth noticing that the three gates are their own inverses. A second application of any of them will undo the effect of the first one. 

\subsection{Multiple qubit gates and controlled operations}
Multiple qubit gates describe complex operations that imply more than one qubit. Although arbitrarily complex gates can be defined, most quantum circuits are based on single and two qubit operations, or $n$ qubit gates from families that allow a simple definition, like Quantum Fourier Transform blocks.

Unlike classical multiple bit gates, quantum gates are always reversible and can create entanglement between qubits. Entangling gates, i.e. gates that can entangle independent qubits, play a fundamental role in quantum information. In particular, controlled operations provide an intuitive formulation of qubit interactions. 

A controlled operation, $CU$, applies the quantum gate $U$ on a group of qubits, the target qubits, if another set of qubits, the control qubits, have a particular value. Control qubits do not change during the process. We follow the usual notation that represents a control by a dot if the gate is activated by a $\ket{1}$ state or a blank circle if it is the $\ket{0}$ state that activates $U$. A controlled gate can have multiple controls acting on a general $U$ operation that can involve multiple qubits. A gate with more than one control is only applied if all the conditions are simultaneously met and will act as the identity operator otherwise.

From the whole range of controlled operations, we will deal primarily with two: the CZ and CNOT operations.

\subsubsection{CZ gate}
Controlled Z, CZ, is the controlled version of the Z gate. This gate applies a Z operation on the target qubit when the control is $\1$. It can also be described as a conditional operation that performs a sign shift only when the two qubit state is $\ket{11}$. From that point of view, it is sometimes referred to as the controlled sign, CS, gate.

The effect on a two qubit state $\ket{x}\ket{y}$ from the computational basis can be summed up as
\begin{equation}
\ket{x}\ket{y} \stackrel{CZ}{\longrightarrow} (-1)^{x\cdot y}\ket{x}\ket{y}.
\end{equation}

\begin{figure}[h]
\centering
\includegraphics{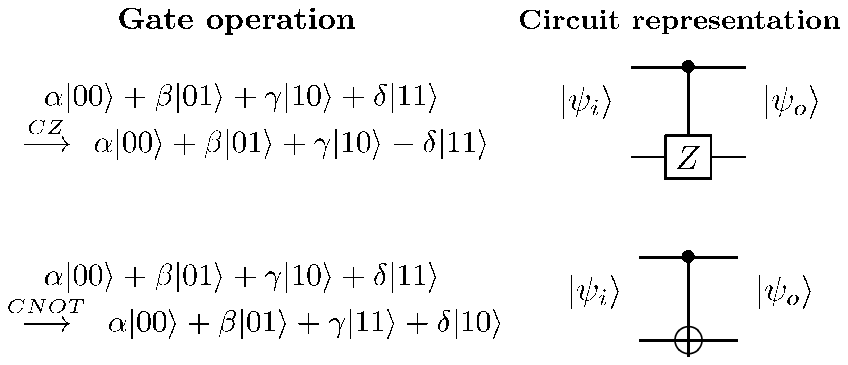}
\caption{Operation of selected controlled gates and their circuit representations.\label{cgatesoperation}} 
\end{figure}

\subsubsection{CNOT gate}  
The most widely used two qubit gate is the CNOT, or CX, gate. It is the controlled version of the X gate and performs a selective negation of the target qubit. The effect of a CNOT gate can be compared to the classical XOR operation. After a CNOT gate, the control qubit is kept while the target now holds the logical XOR of control and target so that
\begin{equation}
CNOT\ket{x}\ket{y}=\ket{x}\ket{x\oplus y}.
\end{equation}

From all the equivalent gates that are essential for quantum computation, the CNOT gate is probably the most well known and it is almost universally used as a fundamental building block of quantum applications. The CNOT gate, when combined with single qubit gates, can provide any desired quantum operation \cite{BBC95}.

Figure \ref{cgatesoperation} shows the usual pictorial representation of the CZ and CNOT gates, along with the description of their effect on a generic pair of qubits that need not to be separable. 

It is sometimes useful to introduce an additional representation of the controlled operations. Truth tables like those employed to illustrate the concepts of classical logic can also be given for quantum gates. Table \ref{Ctruth} presents the truth tables for the CZ and CNOT gates.

\begin{table}[ht]
\begin{center}
\begin{tabular}{ccccccc}	
\multicolumn{3}{c}{\bf{CZ gate}}&&\multicolumn{3}{c}{\bf{CNOT gate}}\\
&&\vspace{-1.5ex}\\
IN& \hspace{1ex}&\phantom{-}OUT&\phantom{aaaa}&IN& \hspace{1ex}&OUT\\
&&\vspace{-2ex}\\
$|00\rangle$&$\rightarrow$&$\phantom{-}|00\rangle$\hspace{1ex}&&$|00\rangle$&$\rightarrow$&$|00\rangle$\hspace{1ex}\\
$|01\rangle$&$\rightarrow$&$\phantom{-}|01\rangle$\hspace{1ex}&&$|01\rangle$&$\rightarrow$&$|01\rangle$\hspace{1ex}\\
$|10\rangle$&$\rightarrow$&$\phantom{-}|10\rangle$\hspace{1ex}&&$|10\rangle$&$\rightarrow$&$|11\rangle$\hspace{1ex}\\
$|11\rangle$&$\rightarrow$&$-|11\rangle$&&$|11\rangle$&$\rightarrow$&$|10\rangle$
\end{tabular}
\caption{Truth tables for the CZ and CNOT gates.}
\label{Ctruth}
\end{center}
\end{table}

Superpositions of input states lead to a superposition of the corresponding output states conserving the associated probability amplitudes.

\subsection{Classically controlled gates}
In some cases, classical and quantum information need to be combined. We can define controlled operations in which the control is a classical bit. We will denote a classically controlled $U$ gate as $cU$. The small c indicates the control is classical instead of quantum.  

\subsubsection{cZ gate} 
The cZ gate will produce a sign shift when the control bit is 1 and the state of the qubit is $\ket{1}$. For a control bit $a$, a state $\ket{x}$ from the computational basis becomes $(-1)^{a\cdot x}\ket{x}$ and
\begin{equation}
cZ(\alpha\ket{0}+\beta\ket{1})=\alpha\ket{0}+(-1)^a\beta\ket{1}.
\label{cZ}
\end{equation}

\subsubsection{cX gate} 
 The cX gate acts in a similar way, but producing a NOT operation instead of a sign shift. For a control bit $b$, a state $\ket{x}$ from the computational basis becomes $\ket{x\oplus b}$ and
\begin{equation}
cX(\alpha\ket{0}+\beta\ket{1})=\alpha\ket{0\oplus b}+\beta\ket{1\oplus b}.
\label{cX}
\end{equation}

\begin{figure}[h]
\centering
\includegraphics{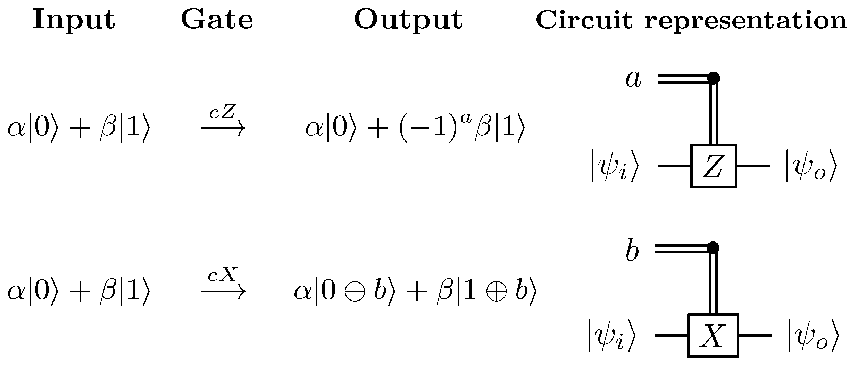}
\caption{Operation of selected classically controlled gates and their circuit representations.\label{ccgatesoperation}} 
\end{figure}

Figure \ref{ccgatesoperation} portrays the most extended notation for classical controlled gates. Classical information is transmitted through classical wires, represented by double lines, while the quantum part is represented as usual. Control is indicated, as in quantum controlled gates, by a dot on the control bit. The dot is connected by double, classical, lines to the controlled gate.

\section{Transformation rules}
\label{rules}

Quantum algorithms and protocols are usually expressed in terms of the quantum circuits that implement them. Quantum circuits are formed by a series of elementary gates that produce the final unitary operation. The sequence of elementary gates is not unique and can be chosen from a number of equivalent circuits.

In this section, we give some basic equivalences that can help us to find simpler physical implementations for a particular application or to design operations when there are particular constraints, usually physically motivated. Additionally, these gate equivalences permit to study the connections between a variety of applications that, on first sight, seem different, but are intimately connected. 

The equivalences will be presented as a series of general rules, followed by the description of some useful cases. The circuits will be mostly composed of H, CZ and CNOT gates and measurements. The point of view will be clearly circuital, with a stress on the usual schematic representation of the gates and circuits. 

\begin{Rule}{Null gates}
\label{nullgates}
Some gates, under certain conditions, have no effect on the qubits they are applied to. Gates that are grouped with their inverses, have a null control or act on the operations' eigenstates with eigenvalue 1, belong to this class.
\end{Rule}

One group of equivalences is the set of the various ways of writing the identity operation. Some gate combinations, or certain gates under particular conditions, are equivalent to the no operation and can be removed from the quantum circuit without a change of the global operation, as long as the conditions are kept. 

Every unitary operation has an inverse. Quantum gates occasionally appear followed by their inverses. A gate $U$ immediately followed by its inverse $U^\dag=U^{-1}$ has no net effect on the input state. Usually, when different functional blocks are chained, these cancellations arise. Separating the elementary operations is essential in the analysis of equivalent circuits. We can add gates that cancel each other to better identify the constitutive blocks of the circuit or use the equivalences to simplify circuits in order to save scarce resources in a physical implementation of a particular operation. 

\begin{figure}[ht!]
\centering
\includegraphics{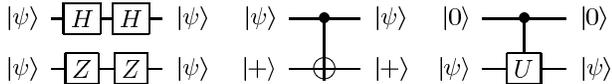}
\caption{Null operations.\label{nullops}} 
\end{figure}

Figure \ref{nullops} shows three kinds of null operation that do not affect the state of the system. The first identity is based on the fact that H, X and Z gates are their own inverses. If any of them is repeated two times in a row, the first one is cancelled by the second. The same can be said of CNOT, CZ and CH gates. 

There is also a CNOT null operation. The $\ket{+}$ and $\ket{-}$ states are the eigenstates for the X operation, with eigenvalues 1 and -1. As a consequence, the $ \ket{+}$ state is not affected by an X gate and, as control qubits do not change, a CNOT with a target $\ket{+}$ state has no effect. 

Finally, controlled gates with a control qubit $\ket{0}$ can also be ignored. In that case, they are not active and are equivalent to the identity operation.

\begin{Rule}{Control reversal}
\label{controlreversal}
In controlled gates, the roles of control and target qubits can sometimes be exchanged. In particular, CNOT gates can be reversed with the help of H gates. 
\end{Rule}

\begin{figure}[ht!]
\centering
\includegraphics{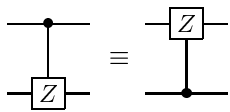}
\caption{Control reversal operation for a CZ gate.\label{CZrev}} 
\end{figure}

On some occasions, the control and target roles are not clear in a controlled gate. Controlled Z gates, CZ, are symmetrical. They induce a sign shift for states where both qubits are $\ket{1}$ and any qubit can be rightfully said to be the control (Figure \ref{CZrev}). In many cases, it can be more illustrative to use an equivalent circuit where control and target roles are exchanged. 

\begin{figure}[ht!]
\centering
\includegraphics{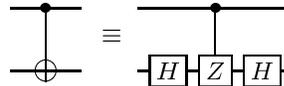}
\caption{CNOT with a CZ gate and two H gates.\label{CNOTH}} 
\end{figure}

The CNOT, or CX, operation can be described in terms of a CZ gate. The X gate can be decomposed as a sequence of three single qubit gates, two Hadamard gates and a Z gate, so that $X=HZH$ and $CNOT=(I\otimes H)CZ(I\otimes H)$ (Figure \ref{CNOTH}). When the control is $\ket{0}$, the two Hadamard gates cancel each other and, when it is $\ket{1}$, the combination of gates acts as a NOT. 

\begin{figure}[ht!]
\centering
\includegraphics{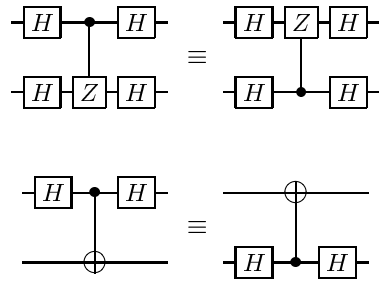}
\caption{CNOT and H gates reversal.\label{CNOTrev}} 
\end{figure}

Figure \ref{CNOTrev} shows how this decomposition of the CNOT gate, together with the control reversal property of the CZ gate, can be used to find equivalent circuits. The control of a CNOT gate can be transferred to the former target when surrounded by the appropriate combination of H gates. 

The starting point is a CZ gate cornered by four Hadamard operations. Grouping the gates in the different qubits, it is easy to see that, in a CNOT with a control sandwiched between two H gates, control and target are interchangeable terms. For the reversal, we consider the gate $(H\otimes I)CNOT(H\otimes I)=(H\otimes I)(I\otimes H)CZ(I\otimes H)(H\otimes I)$. The CZ gate can be reversed and grouped with the upper Hadamard gates to give the lower qubit controlled CNOT. 

\begin{Corollary}
A CNOT gate with four H gates, one before and one after the control and one before and one after the target, is equivalent to a CNOT operation where control and target are exchanged. 
\end{Corollary}

\begin{figure}[ht!]
\centering
\includegraphics{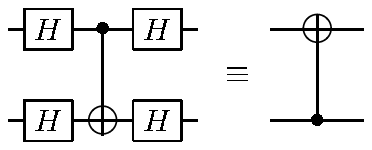}
\caption{CNOT reversal.\label{CNOTflip}} 
\end{figure}

For the proof we only need to add an H gate before and after the target qubit line of the last CNOT circuits of Figure \ref{CNOTrev}. The left side circuit gives a CNOT surrounded by H gates. The right side circuit will present two H gates both before and after the new target. These gates cancel (Rule I).

\begin{Corollary}
A CNOT gate preceded by an H gate in both the control and the target qubits can be reversed moving the H gates after the CNOT.
\end{Corollary}

\begin{figure}[ht!]
\centering
\includegraphics{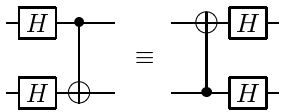}
\caption{H gates mirroring through CNOT inversion.\label{Hmirror}} 
\end{figure}

The equivalence can be deduced from the CNOT circuits of Figure \ref{CNOTflip} adding an H gate after the circuits in both qubit lines. The sequence of two H gates will cancel giving the desired circuits.

\begin{Rule}{Principle of deferred measurement}
\label{defmeas}
A measurement in a qubit line followed by classically controlled operations on other qubits which are controlled by the results of this measurement is equivalent to the corresponding quantum controlled gates with a measurement at the end of the line \em{\cite{NC00}}.
\end{Rule}

\begin{figure}[ht!]
\centering
\includegraphics{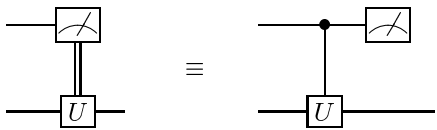}
\caption{Measurement commutes with controlled gates.\label{deferred}} 
\end{figure}

Measurement can commute with controls for certain operations. We will take advantage of this fact later in order to reduce complex quantum gates to simple single qubit operations completed with measurement and classical processing. Figure \ref{deferred} shows an example for the circuit representation. Classical bits are represented with the customary double lines. 

A reduction in the number of purely quantum gates can alleviate the strong constraints that appear in the physical implementation of quantum computers. Quantum states are extremely fragile and must be constantly preserved from decoherence. As long as the function of the whole system is equivalent, classical operations are preferred to classically controlled quantum gates which, in turn, are more desirable than controlled quantum gates. Quantum gates are more prone to error than classical ones. The conversion of quantum domain operations into classically controlled quantum gates, or classical operations, helps to protect the most delicate and critical part of the system, the quantum states.

\begin{Rule}{Quantum-classical substitution}
\label{qcsubs}
Some quantum controlled operations on a pair of qubits which are later measured can be substituted for measurement followed by classical operations. Specifically, a CNOT operation on a qubit which later controls a quantum gate, $U$, can be replaced by the same operation controlled by the classical XOR of the result of measuring the CNOT operands.
\end{Rule}

\begin{figure}[ht!]
\centering
\includegraphics[scale=1]{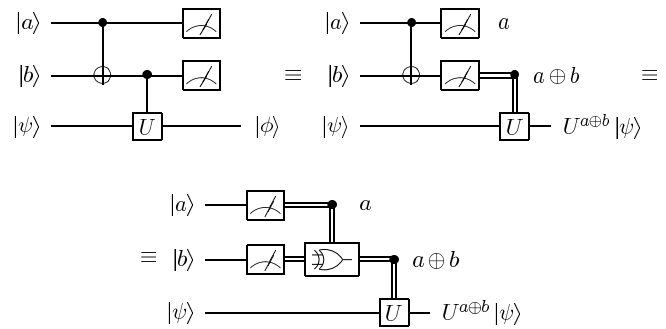}
\caption{Classical substitution of a CNOT gate before measurement.\label{qclassub}} 
\end{figure}

Classical replacement of delicate operations, like in the application the previous rule, can simplify quantum circuit implementation. One further step can be taken if a quantum-classical substitution of gates is applied before a measurement. In our case we will be concerned with the substitution of a CNOT gate for a classical XOR. Imagine a situation like the one depicted in Figure \ref{qclassub}, where there is a controlled quantum gate preceded by a CNOT on two qubits that are going to be measured. The classical XOR gate is represented by the accustomed symbol used in classical circuit schematics. 

The first two qubits will be dubbed as ``control'' and $U$ will taken to be the controlled operation. For input control states in the computational basis, before measuring we have the state $\ket{a}\ket{a\oplus b}$ and the measurement results will be $a$ and $a\oplus b$. The output state is $\ket{\phi}=U^{a\oplus b} \ket{\psi}$. By the principle of deferred measurement, we can convert the CU gate into a classically controlled gate. As the measurement will always yield $a\oplus b$, we can save the CNOT gate and perform the operation classically with a XOR. 

For general control superpositions, we will have general input states of the form $\sum_i \alpha_i \ket{a_i}\ket{b_i}\ket{\psi_i}$ and output states $\sum \alpha_i \ket{a_i}\ket{a_i\oplus b_i}U^{a_i\oplus b_i}\ket{\psi_i}$. Still, after measurement, we will have the same $|\alpha_i|^2$ probabilities of reading $a_i$ and $a_i\oplus b_i$ in the circuit with the CNOT and of finding the corresponding $a_i$ and $b_i$ in the new circuit. In both cases, after the controlled $U$, the state of the lower qubit is $U^{a_i\oplus b_i}\ket{\psi_i}$. If we take the XOR of the measured values as the control, the output will be indistinguishable from the case with a CNOT gate (multiple measurements of the same inputs will have the same statistics).

Although this equivalence works particularly well for the CNOT/XOR conversion, the property cannot be extended to all the quantum controlled gates. Such a substitution can only take place whenever there is a classical gate that reproduces the probabilities for the output state. 

The principle of deferred measurement and this quantum-classical gate substitution can simplify many circuits, especially when there are ancillary qubits. Without loss of generality, we can suppose that after the last operation in which a qubit is involved, it is measured. This way, it is possible to convert some of the operations into simpler classical or classically controlled gates.

\begin{Rule}{Distributed CNOT}
\label{distCNOT}
A CNOT operation between two qubits can be implemented with four CNOT operations with an intermediate qubit so that there is no direct interaction between the original qubits. 
\end{Rule}

\begin{figure}[ht!]
\centering
\includegraphics{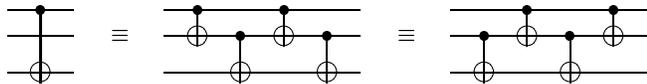}
\caption{Distributed CNOT operation with the intervention of an ancillary qubit.\label{dist}} 
\end{figure}

With this rule, we can redistribute the quantum gates along the different parts of a circuit. The starting point will be the distributed CNOT gate of Figure \ref{dist}. 

The equivalence can be proved from the properties of the XOR function for a particular input and then be generalized to superpositions \cite{Mer01}. For three states $\ket{C}\ket{A}\ket{T}$ of the control, ancillary and target qubit, respectively, the CNOT operation can be expressed as
\begin{equation}
\label{XORCNOT}
\ket{C}\ket{A}\ket{T} \longrightarrow \ket{C}\ket{A}\ket{C \oplus T}.
\end{equation}
For any logical value, $B$, $B\oplus B=0$. We can readily check that the circuits of Figure \ref{dist} recover the final state of Equation (\ref{XORCNOT}). For the middle circuit,
\begin{equation}
\nonumber
\ket{C}\ket{A}\ket{T} \longrightarrow \ket{C}\ket{A\oplus C }\ket{ T} \longrightarrow \ket{C}\ket{A\oplus C }\ket{A\oplus C\oplus T}
\end{equation}
\begin{equation}
\longrightarrow\ket{C}\ket{A}\ket{A\oplus C\oplus T}\longrightarrow \ket{C}\ket{A }\ket{C\oplus T}.
\end{equation}
The second CNOT gate between each pair of lines is there to erase residual correlations. Similarly, for the circuit of the right, 
\begin{equation}
\nonumber
\ket{C}\ket{A}\ket{T} \longrightarrow \ket{C}\ket{A }\ket{A\oplus T} \longrightarrow \ket{C}\ket{A\oplus C }\ket{A\oplus T}
\end{equation}
\begin{equation}
\longrightarrow \ket{C}\ket{A\oplus C}\ket{ C\oplus T}\longrightarrow \ket{C}\ket{A}\ket{C\oplus T}.
\end{equation}

 The new circuit can be useful to implement CNOT gates between distant qubits in cases where only nearest neighbour interactions are possible. With this decomposition of the CNOT gate, we can prove the two next properties related to different ways of arranging CNOT gates.

\begin{Rule}{CNOT mirror}
\label{CNOTmirr}
The order of two chained CNOT gates such that the target qubit of the first is the control of the second can be commuted adding a new CNOT gate from the control of the first CNOT to the target of the second. 
\end{Rule}

\begin{figure}[ht!]
\centering
\includegraphics{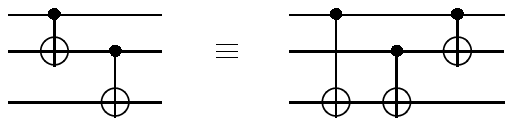}
\caption{Mirror over a CNOT gate.\label{CNOTmirror}} 
\end{figure}

The CNOT mirror operation gives a way to commute CNOT gates acting on different qubits when the control of one of the gates is immediately before or after the target of the other (Figure \ref{CNOTmirror}). The gates can be reflected with respect to a new CNOT gate that has as its control the first gate's control and targets the target of the second. This new CNOT gate commutes with the other two and the reflection can happen on both sides.

\begin{figure}[ht!]
\centering
\includegraphics{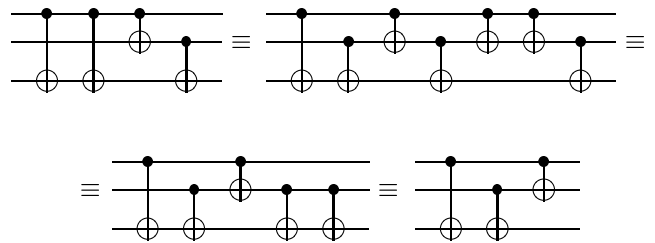}
\caption{Proof of the CNOT mirror.\label{mirrorproof}} 
\end{figure}

To prove mirroring over a CNOT gate, we will use Rules I and V. We can add two CNOT gates before the original gates and then simplify the circuit that arises from the equivalent distributed CNOT of one of the new gates (Figure \ref{mirrorproof}).

\begin{figure}[ht!]
\centering
\includegraphics{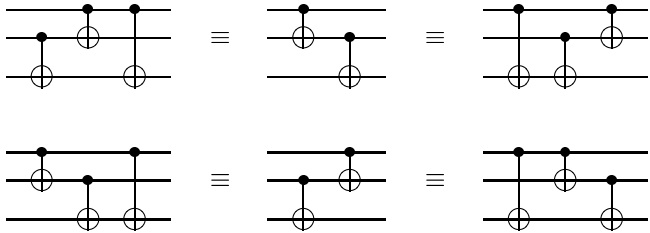}
\caption{Alternative configurations for the mirror CNOT commutation.\label{moremirrors}} 
\end{figure}

Controls commute and so do X gates, so there are many different ways to write the equivalence (see Figure \ref{moremirrors}). In all of them, we can see how the original CNOT gates are ``reflected'' from the longer CNOT and change their order.

These equivalences can be proved with constructions similar to the ones from Figure \ref{mirrorproof}, applying the distributed CNOT decomposition.

\begin{Rule}{Parallel to $\Lambda$ CNOT}
\label{paraCNOT}
Two CNOT gates with a common control qubit and two different targets can be written as three CNOT gates. Two of them are controlled by one of the former targets and act on the other target. The third CNOT gate is placed between them. It conserves the original control and has as the target the control of the new gates.
\end{Rule}

\begin{figure}[ht!]
\centering
\includegraphics{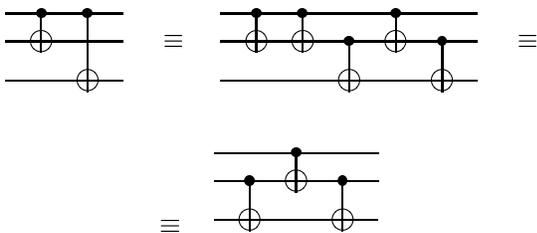}
\caption{Parallel to $\Lambda$ configuration for CNOT gates.\label{Lambda}} 
\end{figure}

When there are parallel CNOT gates with the same control but different targets, they can be rewritten in a $\Lambda$ configuration. Figure \ref{Lambda} shows the resulting circuits from a distributed CNOT equivalence.  

Although the gates commuted when they were in parallel, the $\Lambda$ configuration imposes a fixed order. Controls and X gates only commute under the conditions of the previous rules.

\section{Quantum state transfer}
\label{statetransf}
This section presents a seemingly trivial quantum circuit for quantum state transfer. Following two illuminating papers by David Mermin \cite{Mer01,Mer02}, we will show in Section \ref{examples} that this state transfer is indeed the base for both the teleportation and superdense coding protocols.  

\subsection{Classical XOR swapping and quantum swap circuits}
For two data registers A and B, with content data $a$ and $b$, the problem of state swapping consists in finding a suitable procedure to move the contents of A to B and the contents of B to A. After the transfer, A contains $b$ and B contains $a$. 

There are several different ways to perform this task. One well-known classical swap algorithm is the classical XOR swap algorithm. The algorithm is composed by a sequence of three bitwise XOR operations: 
\begin{enumerate}
\item $(A, A\oplus B)$, 
\item $(A\oplus B',B')$, 
\item $(A',A'\oplus B')$.
\end{enumerate}
For a single bit, it can be summed up as:
\begin{eqnarray}
\nonumber
&&(a,b) \longrightarrow (a,a \oplus b),\\
\nonumber
&&(a,a\oplus b) \longrightarrow (a\oplus a \oplus b,a\oplus b)= (b,a\oplus b),\\
\nonumber
&&(b,a\oplus b) \longrightarrow (b,a\oplus b\oplus b)= (b,a).
\end{eqnarray}

After the three steps the data has been swapped. In all the operations, we can imagine a reversible XOR operation with inputs $(X,Y)$ and outputs $(X,X\oplus Y)$. The algorithm uses reversible logic and, as such, is fit to be extended to quantum information.

\begin{figure}[ht!]
\centering
\includegraphics{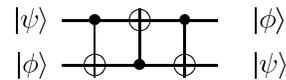}
\caption{Quantum swap circuit.\label{QSWAP}} 
\end{figure}

Substituting the reversible XOR gates by their quantum counterparts, the CNOT gates, we arrive at the quantum circuit of Figure \ref{QSWAP}. 

The swap still takes place for classical data encoded in qubit states from the computational basis in a situation identical to the classical scenario:
\begin{equation}
\nonumber
  \ket{x}\ket{y}\rightarrow \ket{x}\ket{x\oplus y}\rightarrow \ket{x\oplus x \oplus y}\ket{x\oplus y}=\ket{y}\ket{x\oplus y}
\end{equation}
\begin{equation}
\rightarrow \ket{y}\ket{x\oplus y \oplus y}=\ket{y}\ket{x}.
\end{equation}

When the input is a superposition of different computational basis states, the CNOT gate converts each part to its new qubit value with the corresponding probability amplitude. At the output, the input states have been swapped.

\begin{figure}[ht!]
\centering
\includegraphics{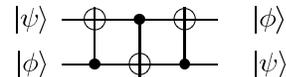}
\caption{Alternative quantum swap circuit.\label{altQSWAP}} 
\end{figure}

The swap operation is highly symmetric. It is completely equivalent to say that A is swapped for B or that B is swapped for A. Both in the classical and the quantum cases, the XOR or CNOT operations can be systematically taken on the opposite register and have the same final result. Figure \ref{altQSWAP} shows an alternative configuration for the quantum swap circuit.

The easiest, and trivial, way to see the circuit performs a swap is noticing that this circuit is exactly the same of Figure \ref{QSWAP}. The lower qubit has been moved to the top and the upper qubit is now down. For the symmetric swap this does not affect the result. 

\section{Bell states}
\label{bellstates}
Many quantum circuits use entangled states as a resource. The most widely used entangled state is the Bell state, or Bell pair
\begin{equation}
\frac{\ket{00}+\ket{11}}{\sqrt{2}},
\end{equation}
a superposition of terms in which both qubits have the same logical value. If we measure each qubit, both outcomes will be the same, even though the particular outcome (0 or 1) is random. This is also true for different measurement bases and for qubits that are taken away a long distance. This kind of correlation cannot be reproduced with classical states \cite{CHS69}.

We can define a family of two qubit Bell states 
\begin{equation}
\label{Belldef}
\ket{\beta_{ab}}=\frac{\ket{0b}+(-1)^{a}\ket{1\bar{b}}}{\sqrt{2}},
\end{equation}
where $\bar{b}$ is $b$ negated, or equivalently, $b\oplus 1$. 

\begin{figure}[ht!]
\centering
\includegraphics{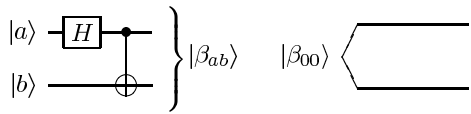}
\caption{Bell state generator (left) and circuit representation of an input Bell pair (right).\label{Bellfig}} 
\end{figure}

The four Bell states are a possible basis of the two qubit Hilbert space. Any two qubit state can be written down as a superposition of $\ket{\beta_{ab}}$ Bell states. Quantum information can be translated from the computational basis into the Bell basis. The Bell state generator from the left of Figure \ref{Bellfig} can take any $\ket{a}\ket{b}$ state from the computational basis into the corresponding Bell pair. This circuit will be an important block when generating Bell states and so will be the matching decoder that results from inverting the gate order to create the inverse operation. 

The term Bell pair, or EPR pair, is usually employed to denote the $\ket{\beta_{00}}$ state, which is represented in quantum circuits with two qubit lines emerging from the same point (see Figure \ref{Bellfig}, right).

\section{Examples}
\label{examples}
\subsection{Quantum teleportation}
\label{teleportation}
One dramatic example of the counterintuitive possibilities of entanglement is the quantum teleportation protocol \cite{BBC93}. In quantum teleportation, the state of an arbitrary quantum system, a single qubit in the basic case, is transferred to another far location with the help of a classical channel and a previously shared Bell pair. 

\begin{figure}[ht!]
\centering
\includegraphics{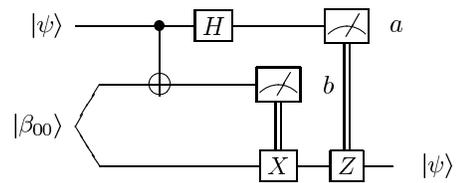}
\caption{Quantum teleportation circuit.\label{basictelep}} 
\end{figure}

The quantum circuit of Figure \ref{basictelep} implements the teleportation protocol for a single qubit.

The circuit works on two stages. First, a Bell pair is separated into distant locations. Then, a CNOT gate correlates the unknown qubit with the first half of the pair. Entanglement between the Bell qubits ensures that there is also a correlation with the remote qubit. The original data needs to be destroyed before teleportation. In order to do that, the first half of the EPR pair and the qubit are measured. Before measurement, the qubit is applied an H gate step that destroys distinguishability between different input qubits. After the H gate, no information on the original qubit state can be deduced from the result of the measurement. At the same time, measurement will project the \emph{remote qubit} into a state that is related to the original qubit. 

The procedure is not complete until the measurement results are sent through a classical channel and are used to correct the state of the remote qubit. The need for this correction guarantees that causality is not violated in spite of the apparent faster than light non-local interaction. No information is sent in the state reduction. 

\subsection{Quantum teleportation as a state transfer}
The quantum teleportation circuit can be derived from the state transfer circuit of Figure \ref{altQSWAP}. We are only interested in taking the input qubit $\ket{\psi}$ to another location. We do not care what state we get in place of the original qubit at the sender. We can imagine we have in the lower input a state $\ket{\phi}=\ket{0}$. This way, we can save the first CNOT gate, which has a $\ket{0}$ control qubit (Rule I).

\begin{figure}[ht!]
\centering
\includegraphics[scale=1]{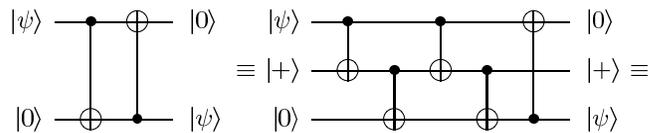}
\caption{Teleportation derived from state transfer.}
\label{tels1}
\end{figure}

Figure \ref{tels1} shows the beginning of the evolution from the simplified state transfer circuit to the most usual teleportation circuit.

The first CNOT can be distributed into four CNOT gates with an ancillary qubit (Rule V). The new qubit has been artificially introduced and its concrete state is not important. We can make it to start in $\+$, which allows to omit the first from the new CNOT gates. 

\begin{figure}[ht!]
\centering
\includegraphics[scale=1]{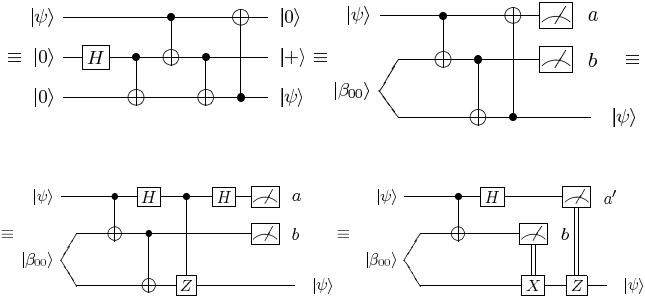}
\caption{Equivalent teleportation circuits.\label{tels2}}
\end{figure}

In Figure \ref{tels2} (up), we substitute the two ancillary qubits in $\+\0$ and the CNOT between them for a Bell pair. The CNOT acting on $\+\0$ is equivalent to a $\0\0$ input for the H and CNOT gates of the Bell state generator of Figure \ref{Bellfig}, which produces $\ket{\beta_{00}}$. The ability to perform this interaction beforehand and then use this Bell pair as a resource gives the teleportation circuit its nonlocal character. 

The lower part of Figure \ref{tels2} depicts more equivalent circuits where these gates have already been replaced by a Bell pair.

Without loss of generality, we can measure the ancillary qubits at the end of the operation. This does not affect the $\ket{\psi}$ state of interest. By replacing the last CNOT gate by two H gates and a CZ gate and reversing this CZ (Rule II), we arrive at the second to last circuit. The last H gate is not important for us and can be omitted, as it only affects the ancillary qubit. Finally, by Rule III, we can replace the last two quantum controlled operations by a cX and a cZ gate and recover the familiar circuit for teleportation.  

\subsection{Dense coding}
\label{densecoding}
Bell pairs can also be employed to send two bits worth of classical information with a single qubit \cite{BW92}. Imagine we have a $\ket{\beta_{00}}$ entangled pair. We can transform this EPR pair into states in the Bell basis with classically controlled gates.

\begin{figure}[ht!]
\centering
\includegraphics{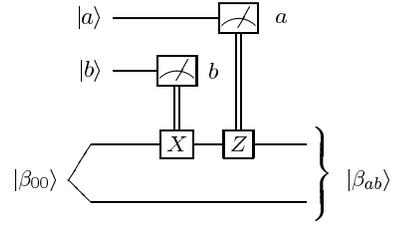}
\caption{Encoding of classical information in a Bell pair using the dense coding protocol.\label{densecirc}} 
\end{figure}

Figure \ref{densecirc} presents a version of this circuit in which the original information is related to the measurement of two qubits in the computational basis that carry the classical information. In this form, its connections to teleportation are clearer. 

The circuit applies a cX and a cZ gate on the first qubit of the Bell pair $\ket{\beta_{00}}$ to take two states $\ket{a}\ket{b}$ from the computational basis into the corresponding state from the Bell basis (the states from Equation \ref{Belldef} up to a global phase). The resulting states can carry the information of two classical bits, but we only needed to act on one of the qubits to encode all the data. This means we can send the second qubit in advance and convey the whole information by just one quantum transmission. The second qubit is not required to have been in the same place as the original information as long as its entangled companion has. 

At the receiver, the inverse of the encoding circuit of Figure \ref{Bellfig}, with a CNOT gate and an H gate, can be used to recover the information in the computational basis. No information can be extracted until both members of the EPR pair are together. As we have seen, entangled states can only be understood as a whole. 

\subsection{Superdense coding as a state transfer}
As it was proved by Mermin \cite{Mer02}, dense coding can be derived from an unremarkable copy circuit that transforms the input state $\ket{x}\ket{y}\ket{0}\ket{0}$ into $\ket{x}\ket{y}\ket{x}\ket{y}$. The copy can be done with two CNOT gates.

The copy circuit is similar to the starting point of the teleportation procedure (Figure \ref{tels1}, left), repeated for each of the qubits. There is a further simplification. As the original states encode classical data, they are only in one state from the computational basis. The last CNOT gate in the state transfer circuit erases residual correlations between the original qubit and the destination qubit. If they were entangled, a measurement on the sender could alter the received state. If the set of possible states is reduced to two orthogonal states, as it happens in dense coding, this last erasure CNOT is no longer necessary.

\begin{figure}[ht!]
\centering
\includegraphics{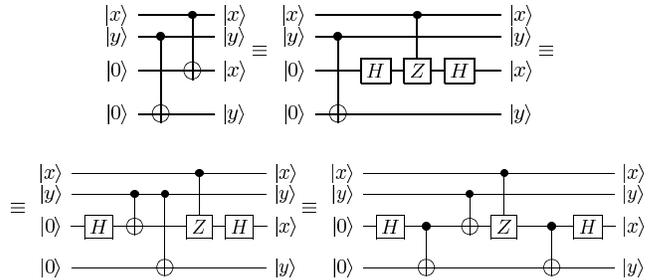}
\caption{Derivation of the superdense coding circuit from a CNOT partial copy of two qubits encoding classical information.\label{dense}} 
\end{figure}

Figure \ref{dense} shows the evolution from the CNOT copying circuit to a superdense encoder. The end circuit has two gates that, when applied to input $\ket{0}\0$, produce the $\ket{\beta_{00}}$ Bell state (see Figure \ref{Bellfig}). It also has the corresponding decoder (a CNOT followed by an H gate) at the receiver's side. 
 
We can use the rules of Section \ref{rules} to show this equivalence. The second CNOT gate of the copying circuit can be written as a sequence of H, CZ and H gates. We can now insert an additional CNOT after the first H gate. The input $\0$ is taken to $\+$, for which, by Rule I, the CNOT gate acts as the identity. Then, by the parallel to $\Lambda$ rule (Rule VII), we can produce two CNOTs between the last qubits and a third one between the second data qubit and the qubit that is going to be transmitted. The further commutation with the CZ gate introduces no changes. The Z operation only introduces a phase shift, but it does not change whether the control is $\0$ or $\1$ and, consequently, whether the NOT is activated or not. 

In the presented circuits, the third qubit is the one that is sent. The fourth qubit can be though to be part of a preshared Bell pair. The data of both bits $x$ and $y$ can be transferred just by acting on the third qubit. The sender only has to apply the CNOT and CZ gates that are controlled by the data to be sent. The counterintuitive data compaction from two bits into one qubit is due to the previously shared entanglement. The strong correlation of the members of a Bell pair allows to treat them as one entity and perform part of the encoding procedure before having the data. 

In this case, we don't even need a complete state transfer circuit. The CNOT copy circuit, applied on two qubits, is enough. Notably, the dense coding protocol cannot be used to send two qubits of information with a single qubit transmission. As we have seen, a complete state transfer circuit for quantum data would require additional entanglement erasure gates (a CNOT for each qubit). Those gates would imply some interaction with the second qubit of the Bell pair (similar to the cZ operation which appeared in teleportation).

\subsection{Gate teleportation}
\begin{figure}[ht!]
\centering
\includegraphics[scale=1]{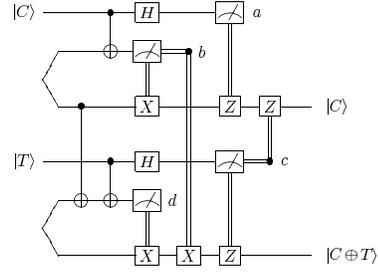}
\caption{Gottesman-Chuang gate teleportation circuit.}
\label{GottChua}
\end{figure}

Gate teleportation offers an alternative way to perform certain operations when there are restrictions that forbid to apply a gate directly \cite{GC99}. Our reference point will be the Gottesman-Chuang gate teleportation circuit of Figure \ref{GottChua}. Here, the input can be put in terms of a new entangled state, $\ket{\chi}=\frac{\ket{0000}+\ket{0011}+\ket{1110}+\ket{1101}}{2}$, which results from the first CNOT operation between the Bell states. With this resource state, we can perform a CNOT operation between remote qubits with classical communication and local classically controlled gates alone. The usual approach to prove this operation is equivalent to a CNOT is based on the commutation rules of the Pauli group operators. We can use a circuital point of view to clarify why this is the case. 

\begin{figure}[ht!]
\centering
\includegraphics[scale=1]{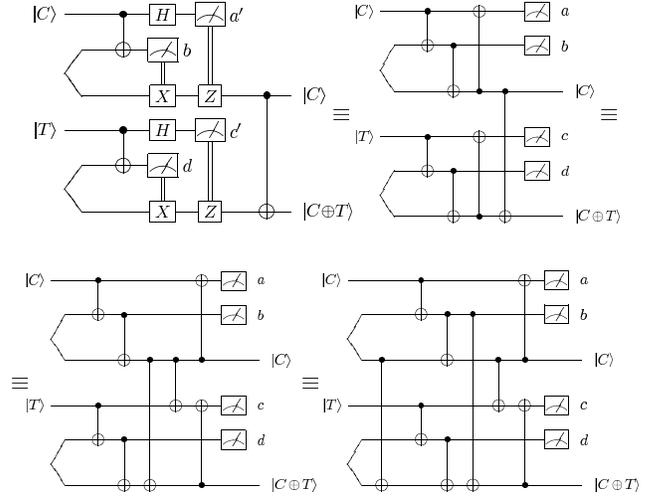}
\caption{CNOT gate teleportation as a teleportation stage followed by a CNOT.}
\label{gatetel}
\end{figure}

We start from two teleportation circuits followed by a CNOT between their outputs (Figure \ref{gatetel}, up). The output is clearly the CNOT operation between control and target. We can replace the standard teleportation circuit with an equivalent circuit with CNOT gates from Figure \ref{tels2}. The measured bits can change, but they are irrelevant for the global operation. 

The last CNOT gate of this circuit can be mirrored in the intermediate CNOTs (rule VI) until it reaches the point of the original Bell pairs (Figure \ref{gatetel}, down). On its way to the beginning of the circuit, the CNOT gate will leave two residual CNOTs which give rise to the crossed cZ and cX gates of the gate teleportation circuit (controlled by bits $b$ and $c$). The final circuit can be readily transformed into the original circuit of Figure \ref{GottChua}. The measurements can be advanced by the principle of deferred measurement. The cZ gates come from the substitution of the CNOT gates by their H and CZ equivalents, as in the usual teleportation circuit (rule II). We ignore two $H$ gates which affect the value of the bits $a$ and $c$, but do not alter the qubits of interest. Finally, due to the symmetry of the Bell pairs, we can see that the CNOT between them has the same effect when it is applied on either qubit of the second pair. The resulting entangled state will be $\ket{\chi}$ in both cases. 

\section{Discussion}
We have presented a series of basic quantum circuit equivalences which can be used to analyse quantum protocols and algorithms. The techniques have been applied to break down the teleportation and superdense coding protocols and to clarify gate teleportation. 

The given equivalences can be a useful analysis tool when dealing with quantum circuits. Transforming the circuits to a different configuration helps to understand the function of each of the elements and the interactions between the different constituting blocks. 

Casting a circuit into a new equivalent form can also be useful in the design of experimental quantum circuits. Each experimental realization of quantum computing has its own strong and weak points. Some gates are easier to implement than others. Searching for an equivalent circuit which reduces the number of problematic gates can improve the final system.

One example are quantum computers implemented with spin chains, where nearest neighbour interactions are well studied and easier to control than long distance interactions \cite{Bos03}. Another area in which the transformation rules can be particularly useful is the optical implementation of quantum computing. One qubit optical gates are easily realizable and there are various options to add a classical control \cite{PJF02a}. By contrast, the construction of CNOT gates has demonstrated to be exceedingly elusive, as photons do not naturally interact by themselves. Using circuit equivalences, the CNOT gates can sometimes be replaced by measurement followed by classically controlled gates \cite{GC09conf,GC09arx}. 

Be it for analysis or design, the equivalences give a quick way to find alternative quantum circuits and gain a fresh point of view. We hope the given compendium will be a valuable addition to the basic toolbox of any quantum computer scientist.

\section*{Acknowledgements}
This work has been funded by projects VA342B11-2 (Junta de Castilla y Le\'on) and TEC2010-21303-C04-04 (MICINN). The figures in this paper have been created using the Q-circuit \LaTeX package \cite{EF04}.

\newcommand{\noopsort}[1]{} \newcommand{\printfirst}[2]{#1}
  \newcommand{\singleletter}[1]{#1} \newcommand{\switchargs}[2]{#2#1}

\end{document}